\documentclass[aps,twocolumn,superscriptaddress,preprintnumbers,showpacs]{revtex4}
\usepackage[colorlinks,linkcolor=blue,urlcolor=blue,citecolor=blue,bookmarks,bookmarksnumbered]{hyperref}
\usepackage{bm}
\usepackage{dcolumn}
\usepackage{graphicx}
\usepackage{booktabs}
\usepackage{multirow}
\usepackage{setspace}
\usepackage{natbib}
\usepackage{cases}

\newcommand{\e}{{\mathrm{e}}}
\renewcommand{\i}{{\mathrm{i}}}

\begin{document}

\newcommand*{\PKU}{School of Physics and State Key Laboratory of Nuclear Physics and
Technology, Peking University, Beijing 100871,
China}\affiliation{\PKU}
\newcommand*{\CIC}{Collaborative Innovation Center of Quantum Matter, Beijing, China}\affiliation{\CIC}
\newcommand*{\CHEP}{Center for High Energy Physics, Peking University, Beijing 100871, China}\affiliation{\CHEP}

\title{Constraints on absolute neutrino Majorana mass from current data}

\author{Yanqi Huang}\affiliation{\PKU}
\author{Bo-Qiang Ma}
\email{mabq@pku.edu.cn}\affiliation{\PKU}\affiliation{\CIC}\affiliation{\CHEP}

\begin{abstract}
 We present new constraints on the neutrino Majorana masses from the current data of neutrinoless double beta decay and neutrino flavour mixing. With the latest results of $0\nu\beta\beta$ progresses from various isotopes, including the recent calculations of the nuclear matrix elements, we find that the strongest constraint of the effective Majorana neutrino mass is from the $^{136}\rm{Xe}$ data of the EXO-200 and KamLAND-Zen collaborations. Further more, by combining the $0\nu\beta\beta$ experimental data with the neutrino mixing parameters from new analyses, we get the mass upper limits of neutrino mass eigenstates and flavour eigenstates and suggest several relations among these neutrino masses.

\end{abstract}


\maketitle

\section{Introduction}
The experiments with solar, atmospheric, reactor and accelerator neutrinos in recent decades~\cite{e1,e2,e3,e4,e5,e6,e7,e8,e9} have provided compelling evidences for the neutrino oscillation. In the three-generation neutrino framework, the neutrino oscillation is caused by the nonzero neutrino masses due to the mixing among the mass eigenstates in the flavour eigenstates.
There have been a number of new measurements on the neutrino mass splitting recently~\cite{MINOS,Daya Bay,T2K new,zhang}. However, the absolute mass scale of neutrinos is still unclear yet. It is also not clarified whether neutrinos are Dirac or Majorana fermions.

Fortunately, the neutrinoless double beta decay ($0\nu\beta\beta$) process provides us some extra information on neutrino properties. The double beta decay of eleven nuclei, whose Q-values are larger than 2~MeV such as $^{76}{\rm Ge}$, $^{82}\rm{Se}$, $^{100}\rm{Mo}$, $^{130}\rm{Te}$ and $^{136}\rm{Xe}$, have been studied up to date~\cite{H-M2001,H-M,CUORICINO,KamLAND-Zen:2012,KamLAND-Zen:2013,KamLAND-Zen:2014,EXO-200,GERDA,EXO-200 new}. For almost all the eleven isotopes, the two-neutrino double beta decay ($2\nu\beta\beta$) processes
\begin{equation}
(A,Z)\rightarrow(A,Z+2)+2e^-+2\bar\nu_e,
\label{decay1}
\end{equation}
are observed. The $0\nu\beta\beta$ process
\begin{equation}
(A,Z)\rightarrow(A,Z+2)+2e^-,
\label{decay2}
\end{equation}
if occurs, can serve as signals of new physics such as the total lepton number violation and the Majorana nature of neutrinos. The only observation of decay events was reported in the year of 2001 by the Heidelberg-Moscow experiment~\cite{H-M2001,H-M}, which claimed the half-life time $T^{0\nu}_{1/2}=2.23^{+0.44}_{-.0.31}\times10^{25} ~{\rm yr}$ on $^{76}{\rm Ge}$ at $68\% \ {\rm CL}$. But the data from the GERmanium Detector Array (GERDA) experiment~\cite{GERDA} in 2013
indicate that the half-life time of $^{76}{\rm Ge}$ should be no lower than $2.5\times10^{25}$ yr (90\% CL), with the Heidelberg-Moscow result unconfirmed. Recently, the Enriched Xenon Observatory (EXO-200) collaboration~\cite{EXO-200 new} and the Kamioka Liquid Scintillator Anti-Neutrino Detector-Zero Neutrino Double Beta (KamLAND-Zen) collabaration~\cite{KamLAND-Zen:2014} released their new results from which we can obtain the strongest constraint on neutrino Majorana mass.
The lower limits on $0\nu\beta\beta$ half time of several isotopes reported by recent experiments are listed in Table~\ref{tab:1}.

We can only obtain the effective Majorana neutrino mass from the half-life time of $0\nu\beta\beta$ process. In order to get constraints of each mass eigenstates and flavour eigenstates, it is necessary to combine the data of double beta decay experiments and oscillation experiments, since the oscillation provides us two mass square differences, as well as the mixing angles $\theta_{12}$, $\theta_{23}$, $\theta_{13}$ and the CP phase $\delta$ in the Pontecorvo-Maki-Nakawaga-Sakata~(PMNS) mixing matrix~\cite{PMNS1,PMNS2}.

In this paper we provide new constraints on absolute neutrino Majorana mass from analysis of the latest data of $0\nu\beta\beta$ processes, combined with also new results of neutrino mixing parameters. In Sec.~II, we obtain the upper limits of the effective Majorana neutrino mass. Then we calculate the mass constraints of different eigenstates by using the CP phase as a variable in Sec.~III and Sec.~IV. In Sec.~V, we present the conclusion and some discussions on some possible relations among neutrino masses.

\section{Neutrino effective Majorana mass}
The half-life time of $o\nu\beta\beta$ process is determined by three kinds of contributions, i.e., the particle mass factor, the phase space integral factor $G^{0\nu}$ and the nuclear matrix element~(NME) $\mathcal{M}_{0\nu}$. Both the latter two vary with different isotopes. If only considering the contribution of the light Majorana neutrino, the half-life time of a given isotope $(A,Z)$ is written as~\cite{formula1,formula2}
\begin{equation}
[T_{0\nu}(0^+\rightarrow 0^+)]^{-1}=\left(\frac{m^{0\nu}_{\beta\beta}}{m_e}\right)^2G_{01}(Q_{\beta\beta}, Z)|\mathcal{M}_{0\nu}(A)|^2,
\label{half-life}
\end{equation}
where $m_e$ is the rest mass of electron. Here we introduce the effective Majorana neutrino mass
\begin{equation}
m_{\beta\beta}^{0\nu}=|\sum_jU^2_{ej}m_j|,
\label{effective mass def.}
\end{equation}
to represent the contribution of the Majorana neutrino during the decay process. $U_{lj}$ ($l=e, \mu, \tau$, $j=1, 2, 3$) are the elements of product matrix of the PMNS matrix $V$ and a diagonal matrix which contains two additional Majorana CP phases $\alpha_2$ and $\alpha_3$,
\begin{widetext}
\begin{equation}
V(\theta_{12},\theta_{23},\theta_{13},\delta)=
\left(\begin{array}{ccc}
            c_{12}c_{13} & s_{12}c_{13} & s_{13}\e^{-\i\delta}\\
           -c_{12}s_{23}s_{13}\e^{\i\delta}-s_{12}c_{23}& -s_{12}s_{23}s_{13}\e^{\i\delta}+c_{12}c_{13}& s_{23}c_{13}\\
             -c_{12}c_{23}s_{13}\e^{\i\delta}+s_{12}s_{23}& -s_{12}c_{23}s_{13}\e^{\i\delta}-c_{12}s_{23}& c_{23}c_{13}\\

      \end{array}\right),
      \label{PMNS}
      \end{equation}

\end{widetext}

\begin{equation}
U=V\cdot \left(
  \begin{array}{ccc}
  1&0&0\\
  0&\e^{\i\frac{\alpha_{21}}{2}}&0\\
  0&0&\e^{\i\frac{\alpha_{31}}{2}}\\
  \end{array}
  \right),
  \label{PMNS Majorana}
  \end{equation}
where $s_{ij}$ and $c_{ij}$ denote $\sin \theta_{ij}$ and $\cos \theta_{ij}$.
The flavour mixing of the three neutrino generations is written as
\begin{equation}
\nu_{l}(x)=\sum_jU_{lj}\nu_j(x),
\label{mixing}
\end{equation}
where the subscripts $l=e,\mu,\tau$ denote the flavour eigenstates and $j=1,2,3$ represent the mass eigenstates.

\begin{table}[h]

\caption{The lower limits of half-life time $T_{1/2}^{0\nu}$ of several isotopes observed by recent experiments~\cite{H-M,CUORICINO,KamLAND-Zen:2012,KamLAND-Zen:2013,KamLAND-Zen:2014,EXO-200,GERDA,EXO-200 new}. }

\begin{ruledtabular}
\label{tab:1}

\centering
\begin{tabular}{ccc}
isotope& experiment&   $T_{1/2}^{0\nu} [{\rm 10^{24} yr}]$ \\

\hline
\noalign{\vspace{0.5ex}}
$^{76}\rm Ge$&	Heidelberg-Moscow&		9.6\\

&	GERDA&	25\\
$^{82}\rm Se$&	NEMO-3&0.32\\
$^{100}\rm Mo$	&NEMO-3&1.0	\\
$^{130}\rm Te$&	CUORICINO&4.1\\
$^{136}\rm Xe$&	KamLAND-Zen 2012&5.7	\\
&KamLAND-Zen 2013&19\\
&KamLAND-Zen 2014&26\\
&	EXO-200	&11\\

\end{tabular}

\end{ruledtabular}

\end{table}

\begin{table}[h]

\caption{The phase space integral factors~\cite{formula2,Q1,Q2} and the nuclear matrix elements from some approximation methods~\cite{ISM,EDF,IBM,QRPA,SRQRPA,SkM-HFB-QRPA} for these five isotopes. }

\begin{ruledtabular}
\label{tab:2}

\centering
\begin{tabular}{cccccc}
isotope &   $^{76}{\rm Ge}$& $^{82}\rm{Se}$& $^{100}\rm{Mo}$& $^{130}\rm{Te}$ & $^{136}\rm{Xe}$  \\

\hline
\noalign{\vspace{0.5ex}}
$G_{01}[\rm{10^{-14}yr^{-1}}]$ &0.63&2.70&4.40&4.10&4.30\\
\hline
\noalign{\vspace{0.5ex}}
ISM(U)&2.81&2.64&-&2.65&2.19\\
EDF(U)&4.60&4.22&5.08&5.13&4.20\\
IBM-2&5.42&4.37&3.73&4.03&3.33\\
QRPA-A&5.16&4.66&5.42&3.90&2.18\\
SPQRP-A&4.75&4.54&4.39&4.16&2.29\\
SPQRPA-B&5.82&5.66&5.15&4.70&3.36\\
SkM-HFB-QRPA&5.09&-&-&1.37&1.89\\

\end{tabular}

\end{ruledtabular}

\end{table}

\begin{table}[]
\caption{The upper limits of effective Majorana neutrino mass from recent experiments, with $m_{\beta\beta,1}^{0\nu}$ and $m_{\beta\beta,2}^{0\nu}$, which are related to the NMEs, denoting the minimal and maximal values of the mass upper limits.}
\begin{ruledtabular}
\label{tab:3}
\centering
\begin{tabular}{cccc}

isotope &    experiment & $m_{\beta\beta,1}^{0\nu}~[{\rm eV}]$&$m_{\beta\beta,2}^{0\nu}~[{\rm eV}]$\\

\hline
\noalign{\vspace{0.5ex}}
$^{76}\rm Ge$	&Heidelberg-Moscow&	0.357&	0.739\\
  &	GERDA&	0.221	&0.458\\
$^{82}\rm Se$ &	NEMO-3&	0.971	&2.082\\
$^{100}\rm Mo$	&NEMO-3	&0.473	&0.923\\
$^{130}\rm Te$&	CUORICINO&	0.243	&0.470\\
$^{136}\rm Xe$&	KamLAND-Zen 2012&0.246&	0.471\\
&KamLAND-Zen 2013&0.135&0.258\\
&KamLAND-Zen 2014&0.115&0.221\\
 &	EXO-200	&0.177&	0.339\\

\end{tabular}

\end{ruledtabular}
\end{table}
\begin{table}[]
      \caption{The global fit of neutrino mixing parameters~\cite{fit2}, with NH and IH denoting the normal and inverted hierarchies of the mass eigenstates. The second square difference of mass is defined as $\Delta m^2=m_3^2-(m_1^2+m_2^2)/2$ in normal hierarchy, and $-\Delta m^2$ in inverted hierarchy.}
      \label{tab:4}
      \begin{ruledtabular}
      \centering
      \begin{tabular}{lll}

       parameter & best fit$\pm 1\sigma$\  & $3\sigma$ range\\
      \hline
      \noalign{\vspace{0.5ex}}
      $\Delta m^2_{21}(10^{-5}~{\rm eV^2})$ & $7.54_{-0.22}^{+0.26}$ & $6.99 \to 8.18$\\
      \noalign{\vspace{0.5ex}}
      $\Delta m^2(10^{-3}~{\rm eV^2})({\rm NH})$ & $2.43_{-0.10}^{+0.06}$ & $2.19 \to 2.62$\\
      \noalign{\vspace{0.5ex}}
      $\Delta m^2(10^{-3}~{\rm eV^2})({\rm IH})$ & $2.42_{-0.09}^{+0.07}$ & $2.17 \to 2.61$\\
      \noalign{\vspace{0.5ex}}
      $\sin^2\theta_{12}({\rm NH\ or\ IH})$ & $0.307_{-0.016}^{+0.018}$ & $2.59 \to 3.59$ \\
      \noalign{\vspace{0.5ex}}
      $\sin^2\theta_{23}({\rm NH})$ & $0.386_{-0.021}^{+0.024}$ & $0.331 \to 0.637$ \\
      \noalign{\vspace{0.5ex}}
      $\sin^2\theta_{23}({\rm IH})$ & $0.392_{-0.022}^{+0.029}$ & $0.335 \to 0.663$ \\
      \noalign{\vspace{0.5ex}}
      $\sin^2\theta_{13}({\rm NH})$ & $0.0241_{-0.0025}^{+0.0025}$ & $0.0169 \to 0.0313$ \\
      \noalign{\vspace{0.5ex}}
      $\sin^2\theta_{13}({\rm IH})$ & $0.0244_{-0.0025}^{+0.0023}$ & $0.0171 \to 0.0313$ \\

      \end{tabular}
      \end{ruledtabular}
\end{table}

For each of the five isotopes mentioned above in Table~\ref{tab:1}, the phase space integral factor is determined and has been calculated~\cite{formula2,Q1,Q2}. However, the NMEs are more complicated and we can only obtain them by approximation methods. There are several common approaches, e.g., the interacting shell model~(ISM)~\cite{ISM}, energy density functional~(EDF) method~\cite{EDF}, interacting boson model (IBM)~\cite{IBM}, quasiparticle random phase approximation~(QRPA)~\cite{QRPA}, self-consistent renormalized quasiparticle random phase approximation~(SRQRPA)~\cite{SRQRPA} and the Skyrme Hartree-Fock-Bogoliubov quasiparticle random phase approximation~(SkM-HFB-QRPA)~\cite{SkM-HFB-QRPA}. The NMEs from some of these methods together with the phase space factors are listed in Table~\ref{tab:2}.

According to Eq.~(\ref{half-life}), we can obtain the expression of the effective Majorana neutrino mass with the half-life time as known quantity,
\begin{equation}
m_{\beta\beta}^{0\nu}=\frac{m_e}{{\mathcal M}_{0\nu}\sqrt{G_{01}T^{0\nu}_{1/2}}}.
\label{effective mass cal.}
\end{equation}
The results of the constrains on $m_{\beta\beta}^{0\nu}$ from different experiments are showed in Table~\ref{tab:3}. Comparing with the half-life time data in Table~\ref{tab:1}, we find that the EXO-200 and KamLAND-Zen experiments provide the strongest constraint on the effective Majorana neutrino mass
\begin{equation}
m_{\beta\beta}^{0\nu}<m_{\beta\beta,\rm min}^{0\nu}=0.115\rm~eV,
\label{effective mass lim.}
\end{equation}
based on the $^{136}{\rm Xe}$ data and EDF approximation approach. It is necessary to notice that though the GERDA group reports a strong lower limit of the $0\nu\beta\beta$ decay half-life time,
it does not provide the strongest constraint on Majorana neutrino mass~\cite{Dev:2013vxa}, since the phase space integral factor of $^{136}{\rm Xe}$ is much larger than that of $^{76}{\rm Ge}$.








\begin{figure}[h]

\begin{minipage}[h] {\linewidth}
\centering      \includegraphics[width=\linewidth]{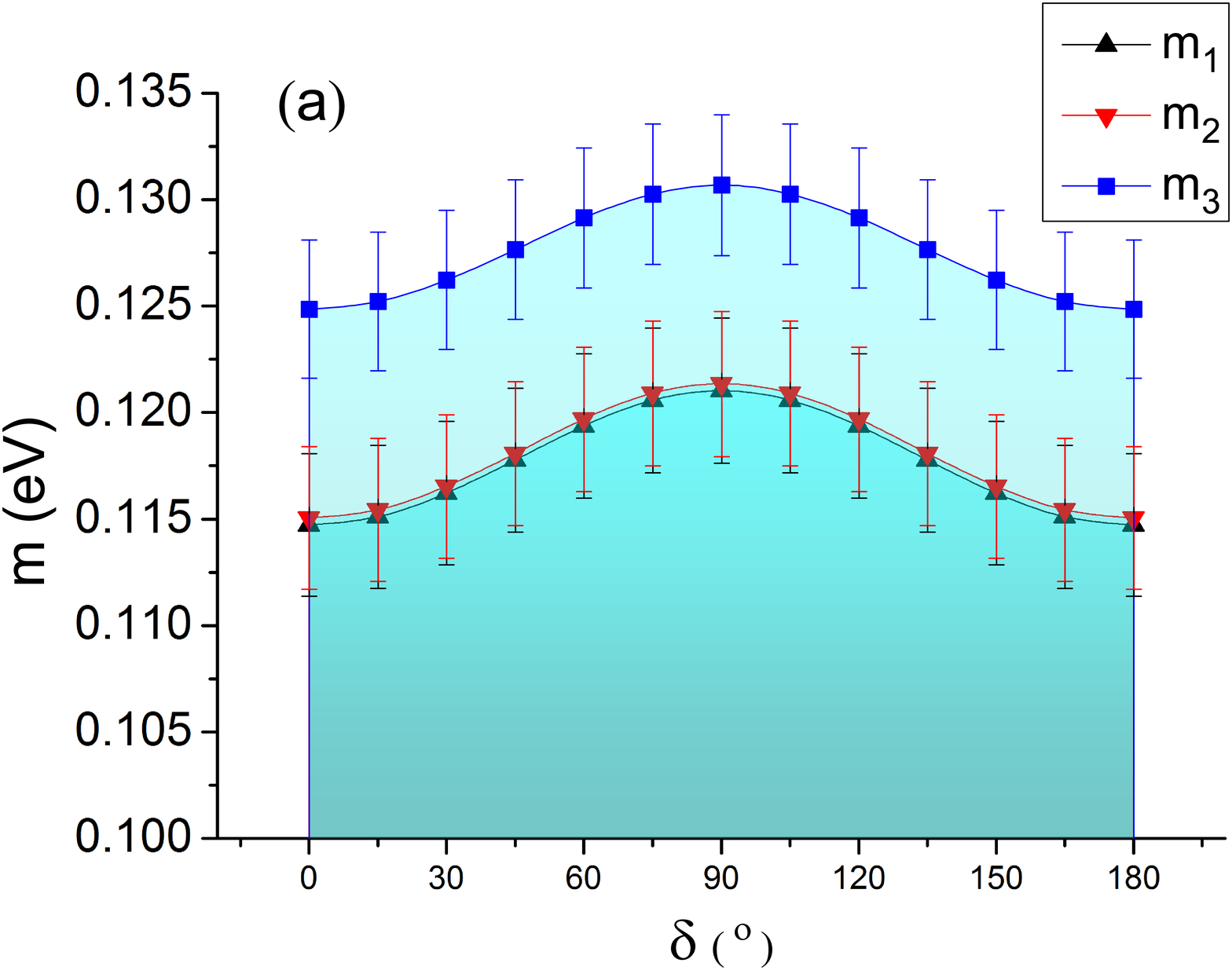}
\end{minipage}

 \begin{minipage}[h]{\linewidth}
 \centering
 \includegraphics[width=\linewidth]{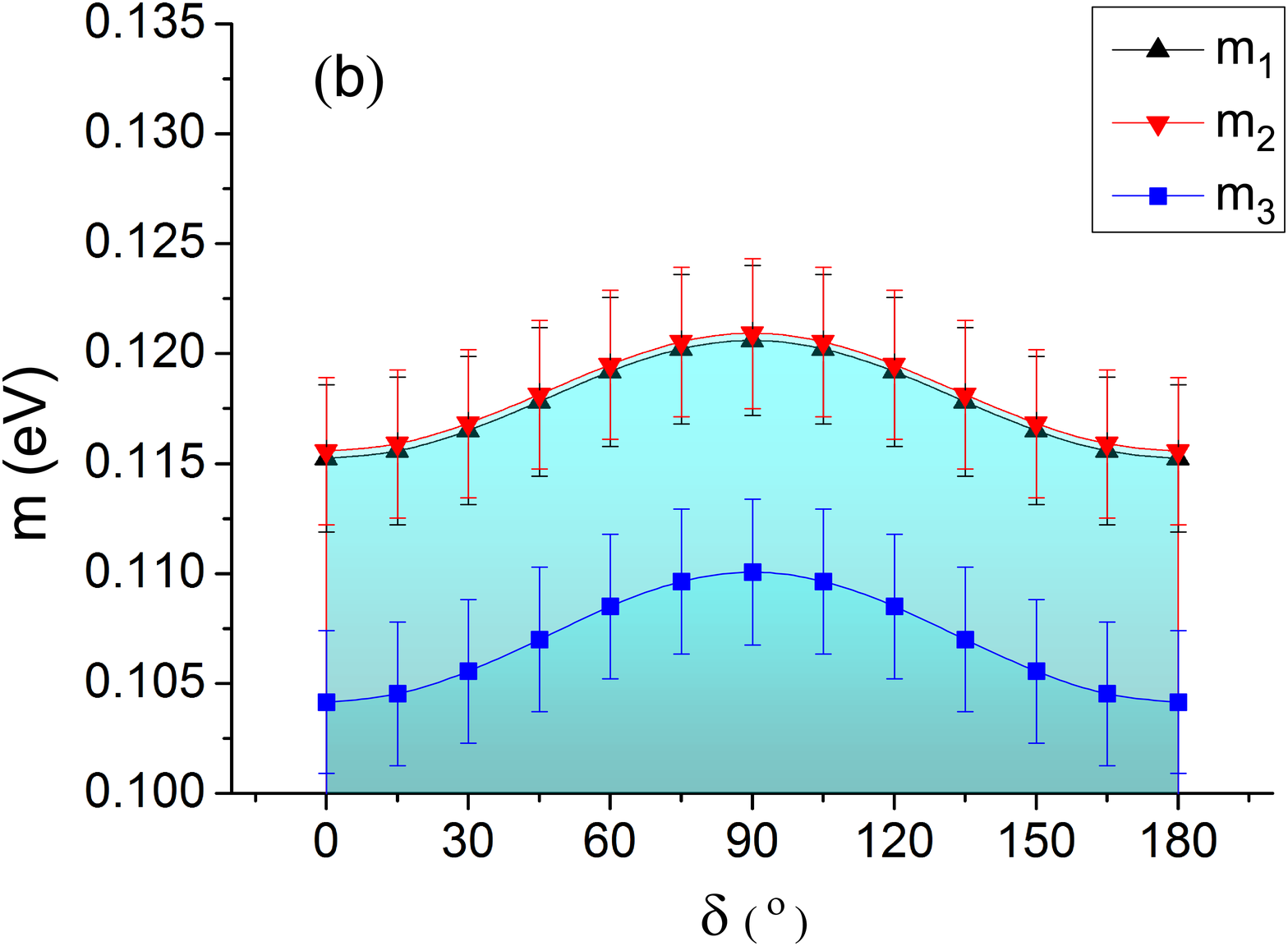}

   \end{minipage}

\caption{The constraints on neutrino Majorana mass eigenvalues, mass upper limits $m_i$ (i=1,2,3) as functions of the CP phase $\delta$. The upper figure (a) corresponds to the normal hierarchy, and the figure (b) for inverted hierarchy. The error bars denote the $\pm 1\sigma$ range influences of the mixing parameters. Since $m_1$ and $m_2$ are very close to each other, the two lines appear to overlap.}
\label{fig:1}
    \end{figure}

\section{Constrains on mass and flavour eigenstates }

Now we already obtain an upper limit of $m_{\beta\beta}^{0\nu}$. According to Eq.~(\ref{effective mass def.}), if considering the mixing parameters we get to
the constraints of three mass eigenstates of Majorana neutrino. Here we use the global fitting results~\cite{fit1,fit2} of mixing parameters listed in Table~\ref{tab:4} as the inputs. The latest T2K result~\cite{T2K} suggests the negative maximal violation of CP phase $\delta=-90^{\circ}$ or $270^{\circ}$, which agrees with the prediction of the maximal CP violation ($\delta =\pm 90^{\circ}$) from the $\mu$-$\tau$ interchange symmetry~\cite{Qu}. In our calculation, we take the CP phase $\delta$ as a parameter too. Since there is little information about the Majorana CP phases, we hypothesize that $\alpha_{21}=\alpha_{31}=0$. Thus, we have only three mass eigenvalues as unknown quantities in the three independent equations below,
\begin{eqnarray}
m_{\beta\beta}^{0\nu}=|(c_{12}c_{13})^2m_1&+&(s_{12}c_{13})^2m_2+s_{13}^2\e^{-2\i\delta}m_3|,~~~~\\
m_2^2-m_1^2&=&\Delta m_{21}^2,\\
m_3^2-m_1^2&=&\Delta m^2+\frac{1}{2}\Delta m_{21}\ (\rm NH),\\
m_1^2-m_3^2&=&\Delta m^2-\frac{1}{2}\Delta m_{21}\ (\rm IH).
\end{eqnarray}
Then we draw the curves that show how the upper limits of neutrino mass eigenvalues $m_1$, $m_2$, $m_3$ change with the CP phase $\delta$, as seen from Fig.~\ref{fig:1}.

In the figure we can find that these three functions $m_i(\delta)$ ($i=1, 2, 3$) are all trigonometric functions with the same phase. The period of the $m_i(\delta)$ is not $2\pi$ but $\pi$.
The $\pi$ period is resulted from the fact that the $\delta$ appears as $2\delta$ form in Eq.~(\ref{effective mass lim.}). If $\delta$ satisfies the maximal CP violation assumption, namely $\delta=\pm 90^{\circ}$, the sign of $\delta$ has no influence for constraining the neutrino Majorana mass in this way, for the range of $\pm 90^{\circ}$ is just the right period of one $\pi$.

 The first two mass eigenvalues $m_1$ and $m_2$ are very close to each other. The third one has a larger difference with respect to them, but the distinction is still within 10~meV. The error bars in the figure reflect the $\pm 1\sigma$ range influences of the mixing parameters. This error range almost covers the difference between $m_1$ and $m_2$. 
 This indicates that the three mass eigenstates $\nu_i$ ($i=1, 2, 3$) are degenerate: $\nu_1$ and $\nu_2$ are strongly degenerate, and $\nu_3$ is a little bit weaker. That helps us to understand why the mixing angle between $\nu_1$ and $\nu_2$ is large. On the other hand, since the level of error is related to $m_{\beta\beta}^{0\nu}$, it is necessary to improve the observation on $0\nu\beta\beta$ process, for getting stronger mass upper limits or even exactly events to enrich the knowledge of the neutrino mass eigenvalues.

 \textbf{\emph{Flavour eigenstates}} --- In the three-generation neutrino framework, there are three mass eigenstates that evolute with the time and three flavour eigenstates that take part in the weak interaction. Equipped with the upper limits of each mass eigenvalue from the effective Majorana neutrino mass, many calculations become feasible. It is the flavour eigenstate mass constraints that interest us most since they have the direct correlation to the dynamic processes. We introduce the mass operator $\hat m$ that satisfies the relation
 \begin{equation}
 m_j=\left<\nu_j|~\hat m~|\nu_j\right>,
 \end{equation}
where $j=1,2,3$ for the mass eigenstates. From Eq.~(\ref{mixing}), the masses of flavour eigenstates, i.e., $m_l=\left<\nu_l|~\hat m~|\nu_l\right>$ ($l=e$, $\mu$, $\tau$), are written as
\begin{eqnarray}
m_e&=&|c_{12}c_{13}|^2m_1+|s_{12}c_{13}|^2m_2+|s_{13}\e^{-\i\delta}|^2m_3,\label{flavour eigenstates1}\\
m_\mu&=&|-c_{12}s_{23}s_{13}\e^{\i\delta}-s_{12}c_{23}|^2m_1\nonumber\\
&+&|-s_{12}s_{23}s_{13}\e^{\i\delta}+c_{12}c_{13}|^2m_2+|s_{23}c_{13}|^2m_3,~~~\label{flavour eigenstates2}\\
m_\tau&=&|-c_{12}c_{23}s_{13}\e^{\i\delta}+s_{12}s_{23}|^2m_1\nonumber\\
&+&|-s_{12}c_{23}s_{13}\e^{\i\delta}-c_{12}s_{23}|^2m_2+|c_{23}c_{13}|^2m_3.~~~\label{flavour eigenstates3}
\end{eqnarray}











\begin{figure}[h]

\begin{minipage}[h] {\linewidth}
\centering      \includegraphics[width=\linewidth]{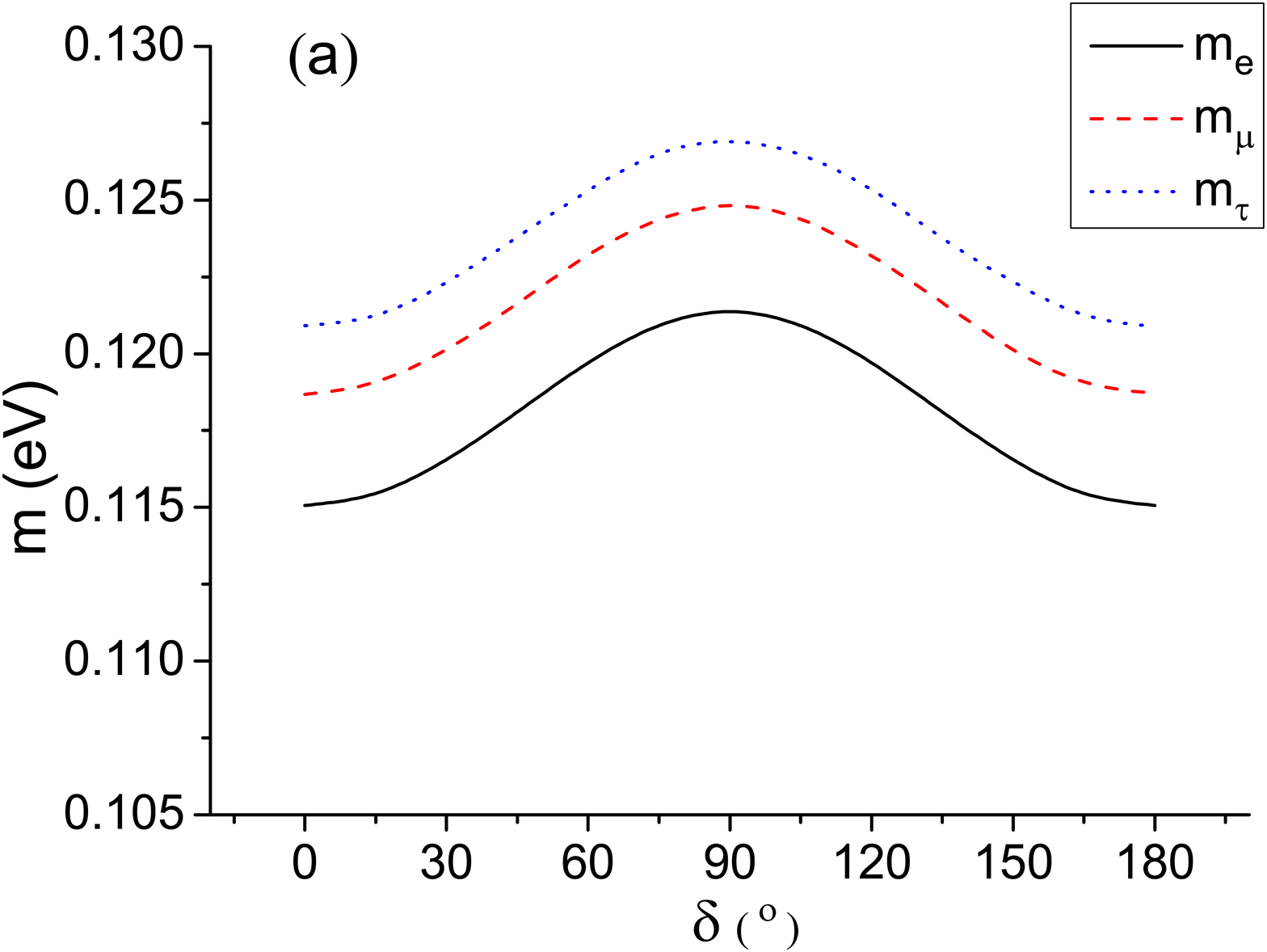}
\end{minipage}

 \begin{minipage}[h]{\linewidth}
 \centering
 \includegraphics[width=\linewidth]{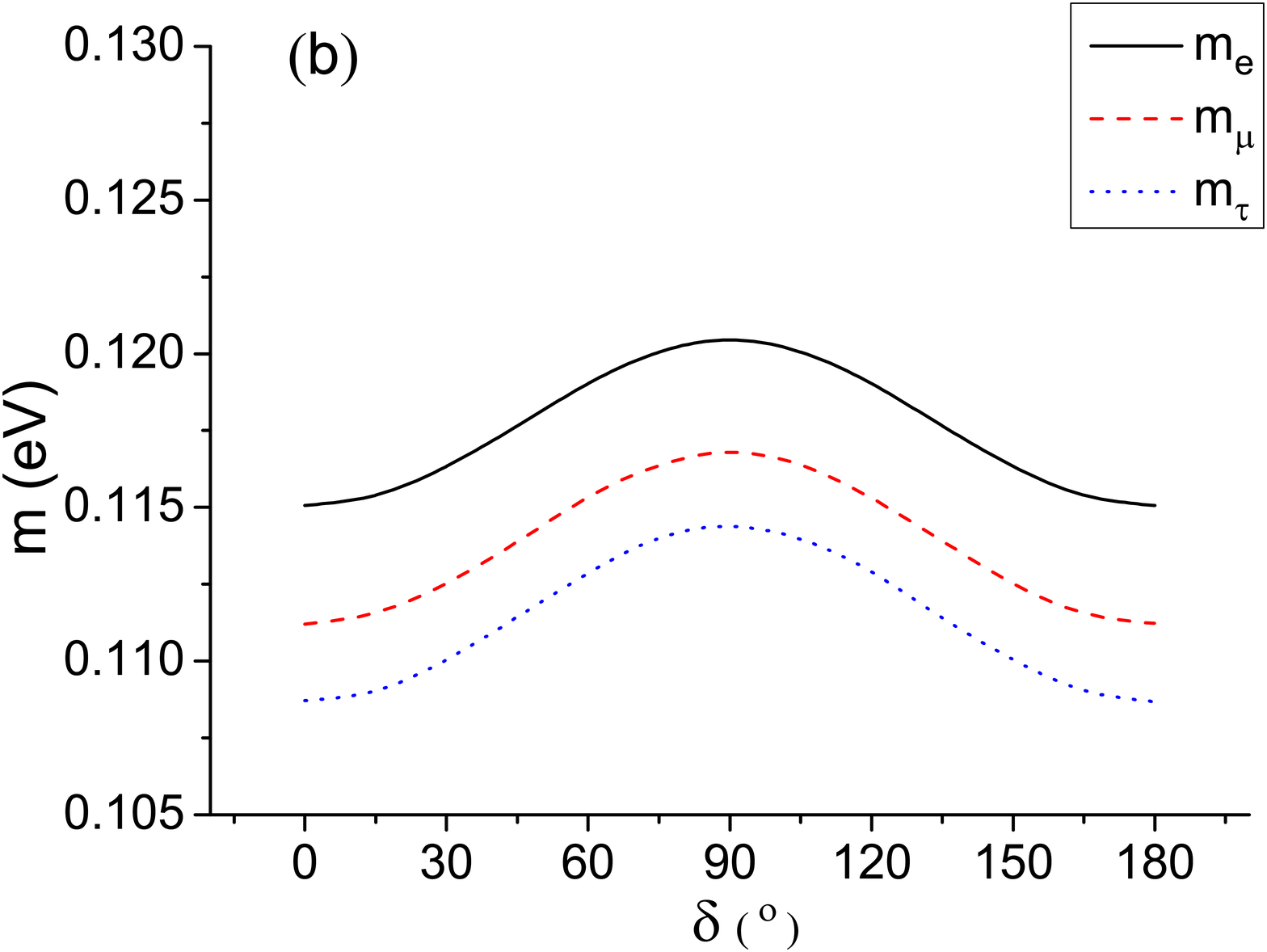}

   \end{minipage}

\caption{The constraints on masses of Majorana neutrino flavour eigenstates. We draw the curves by regarding the CP violate angle $\delta$ as an independent variable, with $m_l$ ($l=e, \mu, \tau$) denoting the mass upper limits. The upper figure (a) corresponds to the normal hierarchy, and the figure (b) for inverted hierarchy.}
\label{fig:2}
    \end{figure}


We use $\delta$ as a parameter to draw the curves of the functions $m_l(\delta)$ ($l=e, \mu, \tau$), as shown in Fig.~\ref{fig:2}. From the figure we find that:
\begin{enumerate}
\item The $m_l$-$\delta$ relations are also trigonometric-like functions. Its frequency is the same as the mass eigenstate. It is easy to understand since $m_i^{\rm lim}(\delta)$ ($i=1, 2, 3$) are trigonometric functions of $\delta$, and the coefficients of $\cos\delta$ terms in Eqs.~(\ref{flavour eigenstates2}) and (\ref{flavour eigenstates3}) cancel each other due to the facts that $\theta_{12} \approx 45^{\circ}$ and
that $m_1^{\rm lim}=m_2^{\rm lim}$.
\item In the NH condition, the mass upper limits of flavour eigenstate masses have the hierarchy
    \begin{equation}
    m_e^{\rm lim}<m_\mu^{\rm lim}<m_\tau^{\rm lim}.\label{hierarchy1}
    \end{equation}
    But in the IH condition, the hierarchy changes into
    \begin{equation}
    m_\tau^{\rm lim}<m_\mu^{\rm lim}<m_e^{\rm lim}.\label{hierarchy2}
    \end{equation}
  Since $m_3$ is the largest (or smallest) one in the NH (or IH) case, Eqs.~(\ref{hierarchy1}) and ({\ref{hierarchy2}}) indicate that the dependence on $m_3$ becomes lager
   when the generation number increases.
    \item We define the differences between mass upper limits of flavour eigenstates as
        \begin{eqnarray}
        \Delta m_{\mu e}&=&m_\mu-m_e, \\
        \Delta m_{\tau\mu}&=&m_\tau-m_\mu.\label{massdif1}
        \end{eqnarray}
        The absolute values of these two differences are the same in the normal hierarchy case and inverted hierarchy case, i.e.,
        \begin{eqnarray}
        \Delta m_{\mu e}({\rm NH})&\approx-\Delta m_{ \mu e}({\rm IH})>0, \\
        \Delta m_{\tau\mu}({\rm NH})&\approx-\Delta m_{\tau\mu}({\rm IH})>0.\label{massdif2}
        \end{eqnarray}
        And $m_e$ is almost the same in the two cases.

\end{enumerate}

From $0\nu\beta\beta$ processes we obtain the upper limit for the summed neutrino mass
\begin{equation}
\sum_{l}m_l=\sum_{j} m_j \le 0.37~{\rm eV}~({\rm NH})~{\rm or}~0.35~{\rm eV}~({\rm IH}),
\end{equation}
which is comparable with the bound 0.23~eV obtained from cosmological observations~\cite{CB}.

\section{Conclusion}
In conclusion, by analysing the latest results of $0\nu\beta\beta$ processes from various isotopes, we give a new constraint of the effective Majorana neutrino mass in ${\rm{^{136}Xe}}$ with the EXO-200 and KamLAND-Zen data~\cite{EXO-200 new,KamLAND-Zen:2014}.
The strongest mass upper limit of $m_{\beta\beta}^{0\nu}$ ranges from 0.115 to 0.339~eV depending on the different approximation methods when calculating nuclear matrix elements. Further more, combining with the global fitting results of neutrino mixing parameters, we calculate the upper mass limits of three mass eigenstates $\nu_1$, $\nu_2$, $\nu_3$ and three flavour eigenstates $\nu_e$, $\nu_\mu$ and $\nu_\tau$. The mass eigenvalue upper limits are very close to each other, and are periodic with period $\pi$. There are some hierarchal and invariant relations which might indicate inner relations among the flavour eigenstates.

The studies on the nature of neutrinos are beneficial to explore new physics. The $0\nu\beta\beta$ decay process is important to explore whether the neutrinos are Majorana fermions or Dirac fermions, and it is also
very useful to provide the limits of their absolute mass scales, as reflected from our analysis. Such kind of experiments can provide more strong constraints on the effective Majorana neutrino mass in the future, and they can
also clarify possible relations among mass eigenstates and flavour eigenstates of neutrinos.

\section{Acknowledgments}

This work is supported by National Natural Science Foundation of China (Grants No.~11035003 and No.~11120101004) and the National Fund for Fostering Talents of Basic Science (Grant Nos.~J1103205 and J1103206). It is also supported by the Principal Fund for Undergraduate Research of Peking University.



\begin{thebibliography}{99}



\bibitem{e1}
 Q.~R.~Ahmad {\it et al.}  [SNO Collaboration],
  Phys.\ Rev.\ Lett.\  {\bf 89}, 011302 (2002)
  [nucl-ex/0204009].

\bibitem{e2}
M.~Altmann {\it et al.}  [GNO Collaboration],
  Phys.\ Lett.\ B {\bf 616}, 174 (2005)
  [hep-ex/0504037].

\bibitem{e3}
S.~Fukuda {\it et al.}  [Super-Kamiokande Collaboration],
  Phys.\ Lett.\ B {\bf 539}, 179 (2002)
  [hep-ex/0205075].

 \bibitem{e4}
  T.~Araki {\it et al.}  [KamLAND Collaboration],
  Phys.\ Rev.\ Lett.\  {\bf 94}, 081801 (2005)
  [hep-ex/0406035].

\bibitem{e5}
 M.~H.~Ahn {\it et al.}  [K2K Collaboration],
  Phys.\ Rev.\ D {\bf 74}, 072003 (2006)
  [hep-ex/0606032].

\bibitem{e6}
P.~Adamson {\it et al.}  [MINOS Collaboration],
  Phys.\ Rev.\ Lett.\  {\bf 101}, 131802 (2008)
  [arXiv:0806.2237 [hep-ex]].

\bibitem{e7}
 P.~Adamson {\it et al.}  [MINOS Collaboration],
  Phys.\ Rev.\ Lett.\  {\bf 107}, 181802 (2011)
  [arXiv:1108.0015 [hep-ex]].



 \bibitem{e8}

  K.~Abe {\it et al.}  [T2K Collaboration],
  Phys.\ Rev.\ Lett.\  {\bf 107}, 041801 (2011)
  [arXiv:1106.2822 [hep-ex]].

  \bibitem{e9}
  F.~P.~An {\it et al.}  [DAYA-BAY Collaboration],
  Phys.\ Rev.\ Lett.\  {\bf 108}, 171803 (2012)
  [arXiv:1203.1669 [hep-ex]].


\bibitem{MINOS}
P.~Adamson {\it et al.}  [MINOS Collaboration],
  Phys.\ Rev.\ Lett.\  {\bf 110}, 251801 (2013)
  [arXiv:1304.6335 [hep-ex]].

\bibitem{Daya Bay}
  F.~P.~An {\it et al.}  [Daya Bay Collaboration],
  Phys.\ Rev.\ Lett.\  {\bf 112}, 061801 (2014)
  [arXiv:1310.6732 [hep-ex]].


\bibitem{T2K new}
  K.~Abe {\it et al.}  [T2K Collaboration],
  Phys.\ Rev.\ Lett.\  {\bf 112}, 181801 (2014)
  [arXiv:1403.1532 [hep-ex]].



\bibitem{zhang}
Y.~Zhang and B.~-Q.~Ma,
  Mod.\ Phys.\ Lett.\ A {\bf 29}, 1450096 (2014)
  [arXiv:1310.4443 [hep-ph]].



 \bibitem{H-M2001}
 H.~V.~Klapdor-Kleingrothaus, A.~Dietz, H.~L.~Harney and I.~V.~Krivosheina,
  Mod.\ Phys.\ Lett.\ A {\bf 16}, 2409 (2001)
  [hep-ph/0201231].


\bibitem{H-M}
H.~V.~Klapdor-Kleingrothaus and I.~V.~Krivosheina,
  Mod.\ Phys.\ Lett.\ A {\bf 21}, 1547 (2006).

\bibitem{CUORICINO}
 E.~Andreotti, C.~Arnaboldi, F.~T.~Avignone, M.~Balata, I.~Bandac, M.~Barucci, J.~W.~Beeman and F.~Bellini {\it et al.},
  Astropart.\ Phys.\  {\bf 34}, 822 (2011)
  [arXiv:1012.3266 [nucl-ex]].

  \bibitem{KamLAND-Zen:2012}
  A.~Gando {\it et al.}  [KamLAND-Zen Collaboration],
  Phys.\ Rev.\ C {\bf 85}, 045504 (2012)
  [arXiv:1201.4664 [hep-ex]].

\bibitem{KamLAND-Zen:2013}
A.~Gando {\it et al.}  [KamLAND-Zen Collaboration],
  Phys.\ Rev.\ Lett.\  {\bf 110}, 
  062502 (2013)
  [arXiv:1211.3863 [hep-ex]].

\bibitem{KamLAND-Zen:2014}
Itaru Shimizu, on behalf of the KamLAND-Zen Collaboration, talk at Neutrino 2104.


\bibitem{EXO-200}
M.~Auger {\it et al.}  [EXO Collaboration],
  Phys.\ Rev.\ Lett.\  {\bf 109}, 032505 (2012)
  [arXiv:1205.5608 [hep-ex]].

\bibitem{GERDA}
M.~Agostini {\it et al.}  [GERDA Collaboration],
  Phys.\ Rev.\ Lett.\  {\bf 111}, 
  122503 (2013)
  [arXiv:1307.4720 [nucl-ex]].

\bibitem{EXO-200 new}
J.~B.~Albert {\it et al.}  [EXO-200 Collaboration],
  Nature {\bf 510}, 229–234 (2014)
  [arXiv:1402.6956 [nucl-ex]].




\bibitem{PMNS1}
 B.~Pontecorvo,
  Sov.\ Phys.\ JETP {\bf 26}, 984 (1968)
  [Zh.\ Eksp.\ Teor.\ Fiz.\  {\bf 53}, 1717 (1967)].

\bibitem{PMNS2}
Z.~Maki, M.~Nakagawa and S.~Sakata,
  Prog.\ Theor.\ Phys.\  {\bf 28}, 870 (1962).


\bibitem{formula1}
 M.~Doi, T.~Kotani and E.~Takasugi,
  Prog.\ Theor.\ Phys.\ Suppl.\  {\bf 83}, 1 (1985).

\bibitem{formula2}
M.~Doi, T.~Kotani and E.~Takasugi,
  Phys.\ Rev.\ C {\bf 37}, 2104 (1988).

\bibitem{Q1}
  M.~Redshaw, E.~Wingfield, J.~McDaniel and E.~G.~Myers,
  Phys.\ Rev.\ Lett.\  {\bf 98}, 053003 (2007).


\bibitem{Q2}
  W.~Rodejohann,
  Int.\ J.\ Mod.\ Phys.\ E {\bf 20}, 1833 (2011)
  [arXiv:1106.1334 [hep-ph]].


\bibitem{ISM}
J.~Menendez, A.~Poves, E.~Caurier and F.~Nowacki,
  Nucl.\ Phys.\ A {\bf 818}, 139 (2009)
  [arXiv:0801.3760 [nucl-th]].

\bibitem{EDF}
T.~R.~Rodriguez and G.~Martinez-Pinedo,
  Phys.\ Rev.\ Lett.\  {\bf 105}, 252503 (2010)
  [arXiv:1008.5260 [nucl-th]].

\bibitem{IBM}
J.~Barea, J.~Kotila and F.~Iachello,
  Phys.\ Rev.\ C {\bf 87}, 014315 (2013)
  [arXiv:1301.4203 [nucl-th]].

\bibitem{QRPA}
F.~Simkovic, V.~Rodin, A.~Faessler and P.~Vogel,
  Phys.\ Rev.\ C {\bf 87}, 
  045501 (2013)
  [arXiv:1302.1509 [nucl-th]].

\bibitem{SRQRPA}
 F.~Simkovic, A.~Faessler, H.~Muther, V.~Rodin and M.~Stauf,
  Phys.\ Rev.\ C {\bf 79}, 055501 (2009)
  [arXiv:0902.0331 [nucl-th]].

\bibitem{SkM-HFB-QRPA}
  M.~T.~Mustonen and J.~Engel,
  Phys.\ Rev.\ C {\bf 87}, 064302 (2013)
  [arXiv:1301.6997 [nucl-th]].

\bibitem{Dev:2013vxa}
  P.~S.~Bhupal Dev, S.~Goswami, M.~Mitra and W.~Rodejohann,
  Phys.\ Rev.\ D {\bf 88}, 091301 (2013)
  [arXiv:1305.0056 [hep-ph]].


\bibitem{fit1}
 M.~C.~Gonzalez-Garcia, M.~Maltoni, J.~Salvado and T.~Schwetz,
  JHEP {\bf 1212}, 123 (2012)
  [arXiv:1209.3023 [hep-ph]].

\bibitem{fit2}
 G.~L.~Fogli, E.~Lisi, A.~Marrone, D.~Montanino, A.~Palazzo and A.~M.~Rotunno,
  Phys.\ Rev.\ D {\bf 86}, 013012 (2012)
  [arXiv:1205.5254 [hep-ph]].

\bibitem{T2K}
 K.~Abe {\it et al.}  [T2K Collaboration],
  Phys.\ Rev.\ Lett.\  {\bf 112}, 061802 (2014)
  [arXiv:1311.4750 [hep-ex]].


\bibitem{Qu}
  H.~Qu and B.~-Q.~Ma,
  Phys.\ Rev.\ D {\bf 88}, 037301 (2013)
  [arXiv:1305.4916 [hep-ph]].


\bibitem{CB}
  P.~A.~R.~Ade {\it et al.}  [Planck Collaboration],
  Astron.\ Astrophys.\  (2014)
  [arXiv:1303.5076 [astro-ph.CO]].

\end{thebibliography}
\end{document}